\documentclass[twocolumn,aps,prl,showpacs]{revtex4}
\usepackage{graphicx}
\usepackage{rotating}           
\def\be{\begin{equation}}
\def\ee{\end{equation}}
\def\bea{\begin{eqnarray}}
\def\eea{\end{eqnarray}}

\def\ra{\rangle}
\def\la{\langle}

\begin{document}
\title{Forecasting non-stationary financial time series through genetic algorithm}
\author{M. B. Porecha$^{1,3}$}
\email{mbporecha@gmail.com}
\author{P. K. Panigrahi$^1$}
\email{prasanta@prl.ernet.in}
\author{J. C. Parikh$^{1}$}
\email{parikh@prl.ernet.in}
\author{C. M. Kishtawal$^2$}
\author{Sujit Basu$^2$}

\address{$^1$Physical Research Laboratory, Ahmedabad, 380 009, India\\
$^2$Space Applications Center, Ahmedabad, 380 015, India\\
$^3$Dharmsinh Desai Institute of Technology, Nadiad, 387 001,
India\\}

\begin{abstract}
We utilize a recently developed genetic algorithm, in conjunction
with discrete wavelets, for carrying out successful forecasts of
the trend in financial time series, that includes the NASDAQ
composite index. Discrete wavelets isolate the local, small scale
variations in these non-stationary time series, after which the
genetic algorithm's predictions are found to be quite accurate.
The power law behavior in Fourier domain reveals an underlying
self-affine dynamical behavior, well captured by the algorithm, in
the form of an analytic equation. Remarkably, the same equation
captures the trend of the Bombay stock exchange composite index
quite well.
\end{abstract}
 \maketitle

It is well-known that a time series, which looks random in nature,
may in fact be the outcome of a nonlinear deterministic but chaotic
dynamics involving a few degrees of freedom. In such cases, it is
possible to exploit this determinism to make short-term forecasts.
Financial time series, originating from complex dynamical processes,
are known to exhibit different types of behavior at different time
scales {\cite{ramsey,mcaul,bisw}}. Random processes like geometric
Brownian motion {\cite{mankiw}} and fractional Brownian motion
{\cite{mandel}} have been invoked for modelling stock market
behavior. Fluctuations in the asset prices have also been analyzed
through Levy-stable non-Gaussian model {\cite{schu}}, a mixture of
Gaussian distributions {\cite{clark}}, etc.  Separation of the
fluctuations at short time scales, owing their origin to random
processes, is essential before attempting any forecast, based on
deterministic dynamics. This is made complicated due to the fact
that, stock market composite indices often show non-stationary
behavior \cite{bouch,kantel}. Wavelets, because of their
multi-resolution capability and time-frequency localization are
ideal to separate out the fluctuations at different time scales in
non-stationary time series \cite{maran}.

The goal of the present paper is to make use of discrete wavelets to
isolate the fluctuations, at different scales, in non-stationary
financial time series data and subsequently employ a recently
developed genetic algorithm for short-term predictions of the trend
\cite{szp,alva,kisht}. We have used NASDAQ and Bombay Stock Exchange
(BSE) composite indices for the purpose of our analysis.

We begin our work with reconstruction of dynamics in phase space
from a time series. Theoretical ideas underlying this reconstruction
are by now well-known and contained in the works of Ruelle
\cite{ruel}, Packard {\it et al.} \cite{packard}, and Takens
\cite{takens}. Thus, given a deterministic time series $x(t_{k}),
t_{k} = k\Delta t, k = 1,..., N$, there exists a smooth map P
satisfying \be \label{geneq} x(t) = P \left[ x(t-\Delta
t),x(t-2\Delta t),...,x(t-\emph{m}\Delta t)\right], \ee where
\emph{m} is called the embedding dimension and $\Delta t$ is the
sampling time interval.

A genetic algorithm tries to obtain the function P, that best
represents the map of a chaotic or integrable time series. The map
can then be used to predict the future state of the system.

{\it Genetic Algorithm.---} The genetic algorithm considers an
initial population of potential solutions consisting of elementary
equation strings. These equation strings (individual solutions) are
of the type as given in Eq.(\ref{geneq}). Their right hand sides are
stored in the computer as sets of character strings that contain
random sequences of the variable at previous times, the four basic
arithmetic symbols (+, -, X, and /), and real number constants. A
criterion that measures how well the equation strings perform on a
training set of the data is its fitness to the data, defined in
Eq.(\ref{rsquare}) in the text. The strongest strings choose a mate
for reproduction whereas the weaker strings become extinct. The
newly generated population is subjected to mutations that change
fractions of information. The evolutionary steps are repeated with
the new generation. The process ends after a number of generation
\emph{apriori} defined by the user.

This technique is first applied to the time series of the NASDAQ
composite index, in the region which shows maximum activity (897
points corresponding to the period between $18^{th}$ February 1999
to $12^{th}$ September 2002). A population consisting of 200 members
was subjected to the iterative algorithm for carrying out one day
ahead forecast. However, it was found that the forecast obtained
from genetic algorithm was only marginally better than persistence
forecast, $x_{t+1}=x_t$. It was thus felt that it would be better if
one attempts to forecast the trend of the data, after removal of
small-scale fluctuations, which owe their origin to random
processes.

\begin{figure}
\label{oritrend}
\includegraphics[width=3in]{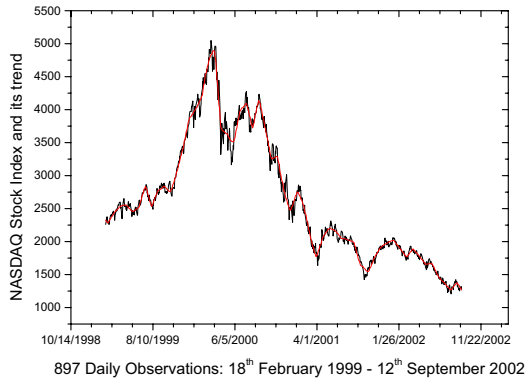}
\caption{(Color Online) NASDAQ composite index (black line) and the
trend extracted through Coiflets (red line) superimposed over it.
The near matching of the two demonstrates the superb ability of the
Coiflets in extracting the trend.}
\end{figure}

{\it Extraction of trend through discrete wavelets.---} For this
purpose, we make use of Coiflets, a family of discrete wavelets,
known to be ideally suited for capturing the trend in a data set
\cite{daub}. Discrete wavelets provide a complete orthonormal basis,
which separates out the average (low-pass) part of a signal from the
variations (high-pass). The orthonormal basis consists of scaling
function (father wavelet) and mother and daughter wavelets. The
scaling function finds out the trend in a data set, whereas the
wavelets identify the variations at different scales. This is
possible since the wavelets have multi-resolution abilities.
Coiflets are compactly supported, with filter length 6N, where N is
the order of the wavelet. These are nearly symmetrical and hence
introduce minimal distortion while capturing the trend. Father
function has 2N-1 vanishing moments, while mother and daughter
wavelet functions have 2N vanishing moments; this endows them with
the property of picking out the trend in an ideal manner.

The time series of the trend was subjected to the genetic algorithm.
The first 700 points were used for training the algorithm and the
remaining points were used for forecast verification. The genetic
evolution process was initiated with 200 randomly selected equation
strings, with \emph{m} set to 4. \emph{m} was found through false
nearest neighbor approach that yielded the value of embedding
dimension as 4 \cite{abarbanel}. It was found that 5000 iterations
were needed for extracting the forecast equation from the set.
Interestingly, the analytic expression involves only three out of
the four input parameters:


\bea \nonumber \label{fit} x_{fit}(t) &=&
P_{fit}\left[x(t-1),x(t-3),x(t-4)\right] \\
&=& \frac{x(t-4)*(x(t-1))^2}{(x(t-3))^2}. \eea

Note that $x_{fit}(t)$ is the prediction of $t^{th}$ day and
$x(t-1),x(t-3)$ and $x(t-4)$ are the trends on the $(t-1),(t-3)$ and
$(t-4)$ days respectively.

The above number of iterations were sufficient in the sense that
increasing the number of iterations did not lead to a significant
increase in the fitness strength:

\be \label{rsquare} R^2 = 1-\left[\frac{\Delta^2}{\sum (x_0 -
\langle x_0 \rangle)^2}\right] \ee

where, $\Delta^2=\sum(x_c-x_0)^2$, $x_c$ is a parameter value
estimated by the best scoring equation, $x_0$ is the corresponding
"true" value, $\la x_0\ra$ is the mean of the "true" values of x.

In Fig.1\ref{oritrend}, we show the time series of the data and the
trend extracted through Coiflets-2, after removal of four level of
high-pass coefficients. Retaining a few dominant high-pass
coefficients, through other thresholding approaches like Donoho, did
not have a significant effect here. More level of decomposition and
thresholding was found to be detrimental to the prediction, since
the same removes significant physical variations from the data set.

\begin{figure}[htb]
\label{trendpred}
\includegraphics[width=3in]{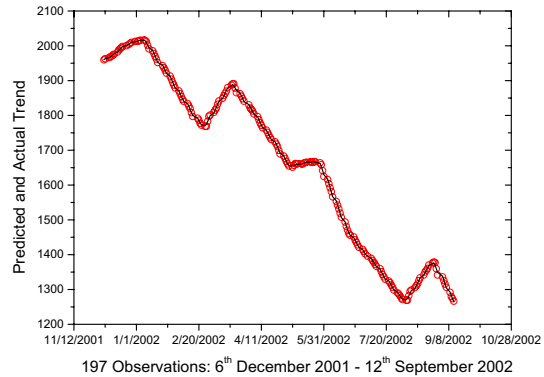}
\caption{(Color Online) A zoomed portion of the NASDAQ composite
index trend superimposed on predicted trend. Here the black line
represents actual trend and the red circle symbol the predicted
trend.}
\end{figure}

Next we made predictions for 747 (out of sample) points which were
not used in training. It is worth pointing out that, since the
embedding dimension \emph{m}=4, we are left with 743 input-output
relations. A zoomed portion of the same is depicted in Fig.2. Fig.3
shows the scatter plot of the predicted and actual trend,
corresponding to time period between $30^{th}$ November 2001 to
$17^{th}$ November 2004, showing a perfect fit. The equation of the
best-fit straight line is, $y = 1.00039 x - 0.59235$ and the
coefficient of determination (square of the coefficient of
correlation) was found to be 0.99994, signifying a high degree of
correlation.

The efficacy of the prediction is further illustrated by the mean
error through the average of modulus of 
return (in percent) which is found to be 0.06831 percent.
Considering the fact that the signature of the trend (rise or fall)
is of significance for financial time series we have further
calculated the following quantity:
$sign(x_{i+1}-x_i)*sign(y_{i+1}-y_i)$ where, $i$ goes from 1 to 742.
The rise and fall are very well captured by the prediction equation
(Eq.(\ref{fit})), since the number of mismatches are only 40 out of
742.

{\it Self-similar dynamics of financial time series.---} We have
analyzed power spectrum of the trend, to ascertain the nature
of this dynamical system and efficacy of the wavelets in removing
small scale fluctuations. It is found that, the power spectrum shows
self-similar behavior; the dynamics have both integrable and chaotic
components. As seen in Fig.4\ref{powertrend}, the power spectrum of
Fourier transform of the trend has a power law decay with exponent
1.88. It has been recently shown that the corresponding exponent for
chaotic dynamics is one, whereas the integrable systems have an
exponent two \cite{rel1,rel2,santh}. Hence, the trend of stock
market dynamics in the present case is closer to an integrable
system.

\begin{figure}[htb]
\label{scatter}
\includegraphics[width=3in]{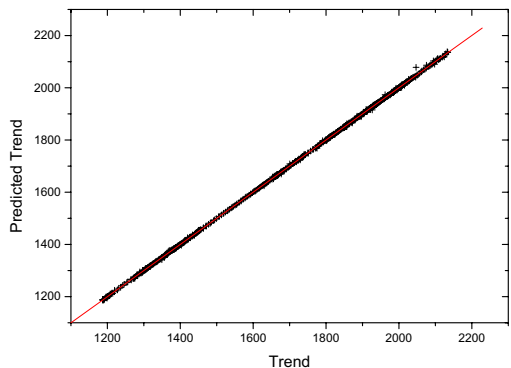}
\caption {(Color Online) The scatter plot of the predicted and
actual trend shows a linear behavior, illustrating the efficacy of
the algorithm. Plus symbol represents scatter plot of the above, red
line represents its linear fitting.}
\end{figure}

\begin{figure}[htb]
\label{powertrend}
\includegraphics[width=3in]{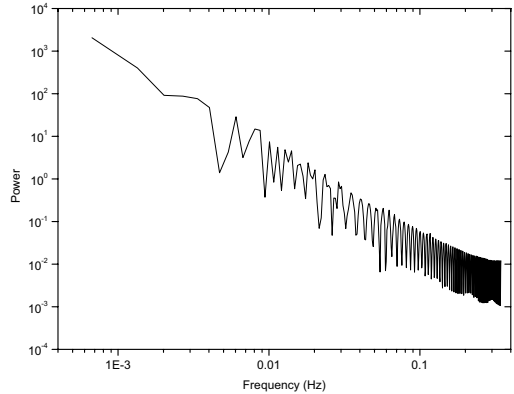}
\caption{Power spectrum of the Fourier transform of the trend, in a
log-log scale. Linear fit ($y = -1.88043 x - 4.26722$) indicates the
closeness of the dynamics to integrable systems, for which the slope
is -2.}
\end{figure}





We have further checked the ability of the present algorithm using
average of daily closing prices of BSE 30 index. This consists of 30
blue chip companies traded on the Bombay Stock Exchange. The
compilation of the values is based on the 'weighted aggregates'
method. It is remarkable that the same equation for the trend as in
Eq.(\ref{fit}) fits the trend of the BSE index, extracted through
Coiflets, extremely well, as seen in Fig.5. This indicates the
underlying similarity between the dynamical processes governing both
the composite indices. The mean error found in prediction of BSE
index was 0.08693 percent.

\begin{figure}
\label{bse}
\includegraphics[width=3in]{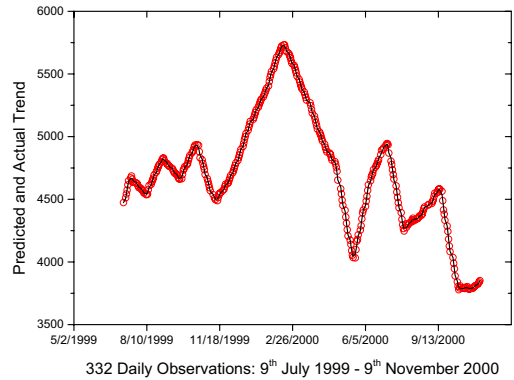}
\caption{(Color Online) A zoomed portion of BSE Sensex trend and its
prediction. Black line represents trend, red circle symbol
represents the predicted trend.}
\end{figure}

In conclusion, a technique has been developed for predicting NASDAQ
and other composite indices' trend using the modern powerful genetic
algorithm and discrete wavelet transform. The algorithm uses the
past values of the trend extracted through wavelets for carrying out
the prediction and is based on the Darwinian theory of survival of
the fittest equation strings. The major advantage of using genetic
algorithm versus other nonlinear forecasting techniques like neural
networks is that an explicit analytic expression for the dynamic
evolution of the trend in the time series is obtained.


It is quite possible that the technique will prove to be successful
in forecasting the trend of individual stocks. The fluctuation in
financial time series also show power law behavior indicating self
similar nature \cite{gaba,chak}, it would be worthwhile to study
their characteristic through this formalism. The fluctuations at
very small scale, will not be amenable for modeling because of their
random origin. Making use of wavelets, one can separate fluctuations
at higher scales for the possibility of prediction. Work in these
directions are in progress and will be reported in future.

The authors are indebted to Dr. A. Alvarez for generously providing
the computer code of genetic algorithm used in the study.


\end{document}